\documentstyle[aps,preprint]{revtex}
\tighten
\begin{document}
\draft
\newcommand{\ie}{{\em i.e.}}
\newcommand{\eg}{{\em e.g.}}
\newcommand{\vs}{{\em vs.}}
\preprint{cond-mat/9412120}
\title{
Magnetization switching in nanoscale ferromagnetic grains:
description by a kinetic Ising model
}
\author{
	 Howard L.\ Richards$^{\rm abc)},$
	     Scott W.\ Sides$^{\rm abc)},$
	     M.\ A.\ Novotny$^{\rm  b)   },$
	and Per Arne Rikvold$^{\rm abc)} $
	}
\address{
  $^{\rm a)}$Center for Materials Research and Technology,
  $^{\rm b)}$Supercomputer Computations Research Institute,
  and
  $^{\rm c)}$Department of Physics,
  Florida State University, Tallahassee, Florida 32306-3016
}

\date{December 23, 1994}
\maketitle
\begin{abstract}
The magnetic relaxation of ferromagnetic powders has been studied
for many years, largely due to its importance to recording
technologies.  However, only recently have
experiments been performed that resolve the
magnetic state of individual sub-micron particles.
Motivated by these experimental developments, we use droplet
theory and Monte Carlo simulations to study the
time and field dependence of some quantities
that can be observed by magnetic force microscopy.
Particular emphasis is placed on the
effects of finite particle size.  The qualitative
agreement between experiments on switching and our simulations in
individual single-domain ferromagnets suggests that the switching
mechanism in such particles may involve local nucleation and
subsequent growth of droplets of the stable phase. 
\vspace{1cm}
\begin{center}
	{\bf FSU-SCRI-94-131}
\end  {center}
\vspace{2cm}
\end{abstract}

\section{Introduction}
\typeout{Introduction}
\label{sec-intro}

Magnetic recording technologies are among the most important
applications of permanent magnets today.
A typical storage medium consists of fine magnetic
particles suspended in an organic binder
adhering to a polymer substrate.
During the recording process, different regions of the medium
are briefly exposed to strong magnetic fields, so that each grain
is magnetized in the desired direction \cite{Koester}.
Each grain could thus in principle store one bit of data,
so greater storage density could ideally be achieved
by a medium containing many small grains than by
one containing a few large grains.
However, in order to serve as reliable storage devices,
the grains must be capable of retaining their magnetizations
for long periods of time in weaker, arbitrarily oriented
ambient magnetic fields. 
Since experiments show the existence of a
particle size at which the magnetizations are most stable
(see, \eg, Ref.~\cite{Kneller63}),
there is a tradeoff between high storage
capacity and long-term data integrity which must give rise
to an optimum choice of grain size for any given material.
During both recording and storage, the relationship between
the magnetic field, the size of the grain, and the lifetime of the
magnetization opposed to the applied magnetic field
is therefore of great technological interest.

Fine ferromagnetic grains have been studied for many
years, but until recently such particles could be
studied experimentally only in powders
(see, \eg, Ref.~\cite{Kneller63}).
This made it difficult to differentiate the statistical
properties of single-grain switching from effects resulting from
distributions in particle sizes, compositions, and local
environments, or from interactions between grains.
Magnetic force microscopy (MFM) (see, \eg,
Refs.~\cite{Martin87,Chang93,Lederman93,Lederman94,Lederm94PRL})
and Lorentz microscopy
(see, \eg, Ref.~\cite{Salling93}) now provide
means for overcoming the difficulties in resolving the magnetic
properties of individual single-domain particles.

The standard theory of magnetization reversal is
a mean-field treatment due to N{\'e}el \cite{Neel49}
and Brown \cite{Brown59,Brown63}.
In order to avoid an energy barrier due to exchange interactions
between atomic moments with unlike orientations,
N{\'e}el-Brown theory assumes uniform rotation of
all the atomic moments in the system.
The remaining barrier is caused by magnetic anisotropy \cite{Jacobs},
either intrinsic or shape-induced. Anisotropy makes it energetically
favorable for each atomic moment to be aligned along one or
more ``easy'' axes.
Buckling, fanning, and curling are, like uniform rotation,
theoretical relaxation processes with few degrees of freedom
and global dynamics \cite{Koester,Kneller}.
In all of the above approaches, the lifetime $\tau$ of the
magnetization increases exponentially with the particle volume,
$L^d$:
\begin{equation}
  \label{eq:MFLife}
	\tau \propto \exp (\beta L^d \Delta f) \ ,
\end  {equation}
where $d$ is the spatial dimension,
$\beta^{-1} \! = \! k_{\rm B}T$ is the temperature in units of energy,
and $\Delta f$ is the height of the free-energy density
barrier to be crossed in the relaxation process.

However, for highly anisotropic materials there exists an
alternative mode of relaxation with a typically much shorter lifetime.
Small regions of the phase in which the magnetization is parallel
to the applied magnetic field (the ``stable'' phase)
are continuously created and destroyed by thermal fluctuations
within the phase in which the magnetization is antiparallel
to the field (the ``metastable'' phase).
As long as such a region (henceforth referred to as a ``droplet'')
is sufficiently small, the short-ranged exchange interaction
with the surrounding metastable phase imposes a net
free-energy penalty, and the droplet will, with high probability,
shrink and vanish.
Should the droplet become larger than a critical size, however,
this penalty will be less than the benefit obtained from orienting
parallel to the magnetic field, and the droplet will with a
high probability grow further, eventually consuming the grain.
The nature of the metastable decay thus depends on the relative
sizes of the grain, the critical droplet, the average distance
between droplets, and the lattice constant,
as discussed in detail, \eg, in
Refs.~\cite{Orihara92,Tomi92A,Rik94,RikARCP94}.

At this point we wish to emphasize the difference between
a droplet and a domain.  Whereas they are both spatially
continuous regions of uniform magnetization, a domain is an
equilibrium feature whereas a droplet is a strictly
non-equilibrium entity.
A droplet may either grow or shrink, but it will not remain
constant.  All of the particles referred to in this paper are
single-domain, which means that in equilibrium their magnetizations
will be uniform, but as we shall see they may nevertheless
decay through the nucleation and growth of droplets.

Because of its simplicity, the kinetic nearest-neighbor Ising
model has been extensively studied as a prototype for
metastable dynamics (see Ref.~\cite{RikARCP94}, and references
cited therein).
In particular, square- and cubic-lattice Ising
systems with periodic boundary conditions have been used to study
grain-size effects in ferroelectric switching \cite{Duiker90,Beale93}.
A related one-dimensional model has been used to
study magnetization reversal in elongated ferromagnetic particles
\cite{Braun93,Braun94c,Braun94a,Braun94b}.
In this article we use the kinetic Ising model to estimate the
switching field, $H_{\rm sw},$ at which magnetization reversal
is thermally induced on experimental timescales for given
temperatures and system sizes.  (Figure~\ref{fig:HsRoad} is a
sketch of switching-field curves
and the four regions of metastable decay for a typical
metastable system with short-range interactions.)
We also find explicit functional forms for what is actually
observed in MFM experiments:
the time- and field-dependent probability
that a grain of given size remains unswitched.

The organization of the remainder of this paper is as follows.
In Sec.~\ref{sec-model} we describe the kinetic Ising model
and the Monte Carlo methods we use to study metastable decay.
A brief summary of the relevant results of
droplet theory is given in Sec.~\ref{sec-droplet}.
We present our simulational estimates of quantities observable
by magnetic force microscopy in Sec.~\ref{sec-simulate}.
In Sec.~\ref{sec-discuss} we conclude with a qualitative comparison
with some recent experiments and discuss some of the
implications of this work.

\section{Model and Numerical Methods}
\typeout{Model and Numerical Methods}
\label{sec-model}

The model is defined by the Hamiltonian
\begin{equation}
	\label{eq-Hamiltonian}
		{\cal H} =
		-J \sum_{\langle i,j \rangle} s_i s_j
		- H L^d m + D L^d m^2 \; ,
\end  {equation}
where $s_i \! = \! \pm 1$ is the $z$-component of the
magnetization of the atom (spin) at site $i,$
$J \! > \! 0$ is the ferromagnetic exchange interaction,
$H$ is the applied magnetic field times the single-spin
magnetic moment.
The sum $\sum_{\langle i,j \rangle}$ runs over all
nearest-neighbor pairs on a square (generally hypercubic)
lattice of side $L$.
In this work we do not study the effects of grain boundaries,
so periodic boundary conditions are imposed.
The dimensionless system magnetization is given by
\begin{equation}
	\label{eq-sysmag}
		m = L^{-d} \sum_{i} s_i \ ,
\end  {equation}
where the sum is over all $L^d$ sites.
The lattice constant is set to unity.

The last term in Eq.~(\ref{eq-Hamiltonian}) allows for an
approximate treatment of the magnetostatic energy. While a
more detailed treatment using magnetic dipole-dipole interactions
would be more physical, this approximation has
a significant advantage in computational efficiency.
For systems with periodic boundary conditions, the first and
third terms of Eq.~(\ref{eq-Hamiltonian}) are
equal when the system size is given by
\begin{equation}
  \label{eq:L_D}
	L_D \approx \frac{2\sigma_\infty(T)}{D} \ ,
\end  {equation}
where $\sigma_\infty (T)$ is the surface tension along a
primitive lattice vector in the limit $L \rightarrow \infty$.
For the two-dimensional Ising model, $\sigma_\infty (T)$
is known exactly \cite{Onsager44}.
The length scale on which we would expect a transition from
single-domain to multi-domain behavior is approximately $L_D$.
In this work we set $D \! = \! 0$, which reduces
Eq.~(\ref{eq-Hamiltonian}) to the standard
Ising Hamiltonian and guarantees that the system will be
single-domain at equilibrium for all system sizes.
Studies of systems with $D \! > \! 0$ will be reported elsewhere
\cite{DGT0}.

The selection of the Ising model is equivalent to requiring
that the anisotropy constant be much greater than
any other microscopic energy scale in the system.
Since a large anisotropy is in fact required for magnetic materials
used in magnetic recording media \cite{Koester},
this idealization may not be too unreasonable.
Simplicity is our main reason for choosing periodic boundary
conditions and a two-dimensional system, particularly since many
equilibrium properties of the two-dimensional Ising model in
zero field are known exactly \cite{Onsager44} and since the
kinetics of metastable decay has been extensively studied
for this model.  As a result, our model systems more closely
resemble ultrathin magnetic films than magnetic grains.
A more realistic simulation of three-dimensional grains
is planned for later study, but we emphasize
that we expect that droplet theory applies to
almost any spin model with high anisotropy.
Accordingly, equations are written in forms appropriate for
arbitrary dimensionality $d$, even though simulations are only
carried out for $d \! = \! 2$.

The relaxation kinetics is simulated by the single-spin-flip
Metropolis dynamic with updates at randomly chosen sites.
This dynamic is realized by the original
Metropolis algorithm \cite{Metro53} and the
$n$-fold way algorithm \cite{Bortz75}.
(For a discussion on the equivalence of the dynamics of these
algorithms, see Ref.~\cite{NovCIP}.)
The acceptance probability in the Metropolis algorithm
for a proposed flip of the spin at site $\alpha$ from $s_\alpha$
to $-s_\alpha$ is defined as
$W(s_\alpha \! \rightarrow \! -s_\alpha)
	\! = \! \min [1, \exp (-\beta\Delta E_\alpha)]$,
where $\Delta E_\alpha$ is the energy change due to the flip.
The $n$-fold way algorithm is similar, but involves the
tabulation of energy classes.  First an energy class is
chosen randomly with the appropriately weighted probability.
A single site is then chosen from within that class with
uniform probability and flipped with probability one.
The number of Metropolis algorithm steps which would be required
to achieve this change is chosen from a geometric probability
distribution \cite{NovCIP}, and the time, measured
in Monte Carlo steps per spin (MCSS), is incremented accordingly.
The $n$-fold way algorithm is more efficient than
the Metropolis algorithm at low temperatures, where the
Metropolis algorithm requires many attempts before a change is
made.

We study the relaxation of the dimensionless system magnetization
starting from an initial state magnetized opposite to the
applied field $(m(t \! = \! 0) \! = \! +1, H \! < \! 0)$.
This approach has often been used in previous studies,
\eg\ in Refs.~\cite{Rik94,Stauffer92}.
For the temperatures employed in this study, the equilibrium
spontaneous magnetizations in zero field are close to unity,
with $0.95 \! < \! m_{\rm s} \! < \! 1$.  Since the applied field
is negative (and generally small), the equilibrium magnetization
is assumed to be approximately $-m_{\rm s}$
and the metastable magnetization is assumed to be given by
$m_{\rm ms} \! \approx \! + m_{\rm s}$.
We use as an operational definition of the lifetime $\tau$
of the metastable phase the mean first-passage time to a
cutoff magnetization $m \! = \! 0$:
\begin{equation}
	\label{eq-define_tau}
		\tau = \langle t(m \! = \! 0) \rangle \ .
\end  {equation}
It has been observed \cite{Rik94} that the qualitative results
discussed below are not sensitive to the cutoff magnetization
as long as it is sufficiently less than $m_{\rm s}$.
Our choice of $m \! = \! 0$ as the cutoff facilitates comparison
with MFM experiments, which are only capable of measuring the
sign of the particle magnetization.
Note that the {\em mean} first-passage time is generally not
the same as the {\em median} first-passage time.

\section{Droplet Theory}
\typeout{Droplet Theory}
\label{sec-droplet}

In this section we briefly review some elements of
droplet theory.  We concentrate specifically on the form of the
lifetime within the four regions (shown in Fig.~\ref{fig:HsRoad})
of metastable decay for a typical
metastable system with short-ranged interactions.
As we will see below [Eqs.~(\ref{eq:CELife}),
(\ref{eq:SDLife}), (\ref{eq:MDLife})],
the exponential dependence of the
metastable lifetime of a system with short-ranged interactions
does not depend exponentially on the system volume, as is
the case for a system with mean-field interactions
[Eq.~(\ref{eq:MFLife})].
For a more complete review, see Ref.~\cite{RikARCP94}.

At zero applied field, the system enjoys true coexistence
between two degenerate equilibrium phases with magnetizations
$m \! \approx \! \pm m_{\rm s} \! \approx \! \pm 1$.
This leads to the identification of a ``Coexistence'' (CE) region
\cite{Tomi92A,Rik94} within which the applied field is so weak
that metastable decay proceeds only through a single,
system-spanning subcritical droplet.
The field limiting the CE region is called the
``Thermodynamic Spinodal''($H_{\rm ThSp}$) \cite{Tomi92A,Rik94}.
The radius of a critical droplet, given by \cite{RikARCP94}
\begin{equation}
  \label{eq:Rc}
	R_c = \frac{(d-1)\sigma_\infty(T)}{2|H| m_{\rm s}} \ ,
\end  {equation}
provides a criterion for estimating the crossover to decay
by a single droplet.
We estimate $|H_{\rm ThSp}|$ by setting $2R_c \! = \! (L-1)$,
which yields
\begin{equation}
	\label{eq:H_THSP}
	|H_{\rm ThSp}| \approx
	\frac{(d-1)\sigma_\infty(T)}{(L-1)m_{\rm s}} \ .
\end  {equation}

For a short-range-force system in the CE region
the free-energy barrier that must be crossed
comes mostly from the creation of an interface at intermediate
system magnetizations.
The resulting form for the lifetime for systems
with periodic boundary conditions is
\cite{Binder81,Binder82,Berg93}
\begin{equation}
  \label{eq:CELife}
	\tau (L,H,T) \approx A(T) \exp
		\left\{
		2 \beta \sigma_L(T) L^{d-1}
		\left[
		1 + O(HL)
		\right]
		\right\} \ ,
\end  {equation}
where $A(T)$ is a non-universal prefactor and $\sigma_L (T)$
is the $L$-dependent surface tension along a primitive lattice
vector.  Any power-law prefactors in Eq.~(\ref{eq:CELife})
are absorbed into this definition of $\sigma_L (T)$ as
$O[(\ln L)/L]$ terms.
The $O(HL)$ term comes from bulk contributions
to the free energy.

The region of fields just stronger than $H_{\rm ThSp}$
is called the ``Single-Droplet'' (SD) region \cite{Tomi92A,Rik94}.
In this region the first critical droplet to nucleate
almost always grows to fill the system before any
other droplet has a chance to nucleate.
The lifetime in the SD region is given in terms of the nucleation
rate per unit volume $I$ by \cite{RikARCP94}
\begin{equation}
  \label{eq:SDLife}
	\tau (L,H,T) \approx \left[ L^d
		I (T,H) \right]^{-1} \ .
\end  {equation}
The nucleation rate $I (T,H)$ is given by
\begin{equation}
  \label{eq:NucRate}
	I (T,H) \approx B(T) |H|^{K} \exp
	     \left\{ - |H|^{1-d}
		\left[ \Xi_0(T) + \Xi_1(T) H^2 \right]
             \right\} \ ,
\end  {equation}
where $B(T)$ is a non-universal prefactor,
and $K$ is believed to be
    3  for the   two-dimensional Ising model and
$-1/3$ for the three-dimensional Ising model \cite{Rik94}.
Here $\Xi_0 (T)$ is given by \cite{RikARCP94}
\begin{equation}
  \label{eq:def-Xi0}
	\Xi_0 (T) \equiv \beta \Omega
		\left[ \sigma_{\infty} (T)
		\right]^d
		\left[
		  \frac{ d-1 } {2m_{\rm s}}
		\right]^{d-1} \ ,
\end  {equation}
where $\Omega$ is a shape- and dimension-dependent constant
such that the volume of a droplet is given in terms of its
radius by $V \! = \! \Omega R^d$.
For the two-dimensional Ising model, $\Omega$
is exactly known \cite{RikARCP94}.
We determine $\Xi_1$ from a numerical fit.

In both the CE and SD regions, switching is abrupt,
with a negligible amount of time being spent in configurations
with magnetizations other than $\pm m_{\rm s}.$
This phenomenon, in which the entire system behaves as though
it were a single magnetic moment, is known as
superparamagnetism \cite{Jacobs,Bean59}.
Switching is a Poisson process in both the CE and SD regions, so
the probability $P(m \! > \! 0)$ that the magnetization is greater
than zero decays exponentially with time, and the median switching
time is larger than the mean switching time by a factor of $\ln 2$.
The standard deviation of the switching time for an individual
grain is approximately equal to the mean switching time, $\tau$.
Because of the random nature of switching in the
CE and SD regions, their union has been called the
``Stochastic Region'' \cite{Tomi92A,Rik94}.

Still stronger fields, by contrast, lead to
decay through many weakly interacting
droplets in the manner described by
Kolmogorov \cite{Kolmogorov37},
Johnson and Mehl \cite{JohnsMehl39}, and
Avrami \cite{Avrami39,Avrami40,Avrami41}.
Such decay is ``deterministic'' in the sense that
the standard deviation of the switching time is
much less than its mean (see the Appendix).
The crossover between the SD region and the multi-droplet (MD)
region has been called the
``Dynamic Spinodal'' (DSp) \cite{Tomi92A,Rik94}.
Since the standard deviation of the lifetime is equal
to its mean in the Stochastic region,
we estimate this crossover by the field $H_{1/2}$ at which
\begin{equation}
	\label{eq:H_DSP}
	\sqrt{ \langle t^2( m \! = \! 0) \rangle - \tau^2 }
		= \frac{\tau}{2}\ .
\end  {equation}
For {\em asymptotically} large $L$, $H_{\rm DSP}$ tends to
zero as \cite{Rik94}
\begin{eqnarray}
	H_{\rm DSP} & \sim &
		\left[
			 \frac{\Xi_0 (T)}{(d+1) \ln L}
		\right]^{1/(d-1)}
		\left\{ 1 +
	\frac{K-1}{d^2-1} \left[ \frac{\ln (\ln L)}{\ln L} \right]
		\right. \nonumber \\
	  \label{eq:H_DSP_Scale} & & \left.
		  + \left[ \ln \left( \frac{\nu}{B} \right)
	- \frac{K-1}{d-1} \ln \left( \frac{\Xi_0}{d+1} \right)
        + \frac{\ln (2\Omega)}{d} \right]
			\left( \frac{1}{(d + 1)\ln L} \right)
		\right\} \, ;
\end  {eqnarray}
however, extremely large system sizes may be required
before this scaling form is observed.

Within the MD region the lifetime is given by
\cite{RikARCP94}
\begin{equation}
	\label{eq:MDLife}
	\tau (L,H,T) \approx
	\left[ \frac{\Omega v^d}{(d+1)\ln 2} I(T,H)
		\right]^{- \frac{1}{d+1}} \ ,
\end  {equation}
where $I (T,H)$ is given by Eq.~(\ref{eq:NucRate}).
Here $v$ is the (nonuniversal) temperature-dependent
radial growth velocity of a droplet, which under an
Allen-Cahn approximation \cite{Lifshitz62,Chan77,Allen79}
is proportional to the applied field in the limit of
large droplets:
\begin{equation}
  \label{eq:def_nu0}
	v \approx \nu |H| \ .
\end  {equation}

At a sufficiently high field, nucleation becomes much faster than
growth and the droplet picture breaks down.
The crossover to this ``Strong-Field'' region (SF) has been
called the ``Mean-Field Spinodal'' (MFSp) \cite{Tomi92A,Rik94}.
A conservative estimate for this crossover field is obtained by
setting $2 R_c \! = \! 1$:
\begin{equation}
	\label{eq:H_MFSP}
	|H_{\rm MFSp}| \approx \frac{(d-1) \sigma_\infty(T)}
			    {m_{\rm s}} \ .
\end  {equation}

\section{Monte Carlo Simulations of MFM Observables}
\typeout{Monte Carlo Simulations of MFM Observables}
\label{sec-simulate}

The droplet-theory concepts of Sec.~\ref{sec-droplet} provide
a framework for the calculation of a number of observable quantities.
In this section we discuss calculations of MFM observables from
Monte Carlo simulations of the two-dimensional
kinetic Ising model as discussed in Sec.~\ref{sec-model}.

The mean first-passage time $\tau$ to $m \! = \! 0$ can be
found at a fixed magnetic field directly from Monte Carlo
simulations.
Finding the magnetic field corresponding to a given lifetime
requires an indirect approach. From four fields chosen such
that $\tau(L,H,T) \! \approx \! t$ at fixed system size $L$
and temperature $T$, we use weighted
linear regression to estimate the switching field
$H_{\rm sw}(L;t,T)$ such that $\tau(L,H_{\rm sw},T) \! = \! t.$
The results, which are shown in
Fig.~\ref{fig:Hs_vs_L} for three different
temperatures and a range of system sizes and waiting times,
are qualitatively similar to experimental measurements of the
same quantity made on powders
(see, \eg, Ref.~\cite{Kneller63})
and on single-domain particles \cite{Chang93}.
As Fig.~\ref{fig:Hs_vs_L} illustrates, the maximum value
of $H_{\rm sw}(L;t,T)$ occurs within the SD region.
The asymptotically nonzero value of $H_{\rm sw}(L;t,T)$
in the MD region is not observed in actual experiments,
perhaps because defects and impurities lead to
heterogenous nucleation in large grains.
Lastly, a comparison of parts (a), (b),
and (c) of Fig.~\ref{fig:Hs_vs_L} shows the striking temperature
dependence of $H_{\rm sw}$ for a fixed waiting time.

{}From MFM experiments one does not measure magnetization,
but rather detects the orientation of the magnetic poles,
\ie, the sign of $m$.
An experimentally measurable quantity that can be used to
determine the decay mode is therefore
the probability that the magnetization is greater
than zero, $P(m \! > \! 0),$ as a function of field at fixed
waiting time or as a function of waiting time at fixed field.
The particular functional form of $P$ which is predicted by
droplet theory depends upon the region
(CE, SD, MD, or SF) in which $P(m \! > \! 0) \! \approx \! 1/2$.

In the CE region, switching is a Poisson process and the
decay of $\langle m(t) \rangle$ has an exponential form.
However, recrossing (returning to an $m \! > \! 0$ phase from
an $m \! \leq \! 0$ phase) is important because the energy
difference between the ``stable'' and ``metastable'' phases
is so small.
This is most easily seen at $H \! = \! 0$, where the two phases
coexist and
\begin{equation}
  \label{eq:P_H0}
  	P(m \! > \! 0) =[\exp (-2t / \tau) + 1]/2 \ .
\end  {equation}
The factor of two in the exponential comes from the fact that
$\tau$ is the mean first-passage time to $m \! = \! 0$, not to
$m \! = \! -m_{\rm s}$. Figure~\ref{fig:P_vs_t}(a) shows
MC estimates of $P(m \! > \! 0)$ as a function of time, and
Fig.~\ref{fig:P_vs_H}(a) shows MC estimates of $P(m \! > \! 0)$
as a function of field in the CE region.

Because the energy difference between the stable and metastable
phases is large in the SD region,
the probability of return to $m \! > \! 0$ from
$m \! \leq \! 0$ is negligible.
As a result, $P(m \! > \! 0)$ decays exponentially in time,
and we specifically obtain
\begin{equation}
   \label{eq:P_SD}
	P(m \! > \! 0) = \exp \left[
		\frac{-t}{\tau(L,H,T)}
		          \right] \ ,
\end  {equation}
where $\tau(L,H,T)$ is given by Eqs.~(\ref{eq:SDLife}) and
(\ref{eq:NucRate}).
Note that $P(m \! > \! 0)\Bigl|_{\textstyle t=\tau} \Bigr.
\! = \! 1/e \! \neq \! 1/2$ because the mean of an exponential
distribution is not the same as its median.
In Fig.~\ref{fig:P_vs_t}(b), Monte Carlo estimates of
$P(m \! > \! 0)$ as a function of $t$
at fixed $H,$ $T,$ and $L$ are seen to agree well with
an exponential time decay.
Figure~\ref{fig:P_vs_H}(b) shows Monte Carlo estimates of
$P$ as a function of $H$ at fixed $t,$ $T,$ and $L$ together with
fits based on Eqs.~(\ref{eq:SDLife}), (\ref{eq:NucRate}),
and (\ref{eq:P_SD}).
The fit shown in Fig.~\ref{fig:P_vs_H}(b) comes from
a least-squares fit used to determine $B(T)$ and $\Xi_1 (T)$
in Eq.~(\ref{eq:NucRate}) \cite{SDBXi1}.
The fit is seen to agree excellently with the MC data.

The form of $P(m \! > \! 0)$ in the MD region is somewhat more
complicated and is derived in the Appendix.
Both the switching field and the lifetime have asymptotically
Gaussian probability distributions.
Expanding in time around $t \! = \! \tau$ we find
\begin{equation}
  \label{eq:P_vs_t_MD}
	P(m>0)\Bigl|_H \Bigr.   \approx   \frac{1}{2} \left[ 1 +
		{\rm erf} \left( \frac{t - \tau}{L^{-d/2} \Delta_t}
		\right) \right] \ ,
\end  {equation}
where $\tau$ is given by Eq.~(\ref{eq:MDLife}) with
$H \! = \! H_{\rm sw}$.
Here $\Delta_t$ is given by
\begin{equation}
  \label{eq:Delta_t}
	\Delta_t = \frac{( \nu |H_{\rm sw}| )^{d/2} \tau^{(d+2)/2}}
			{(d+1)\psi_d}
\end  {equation}
and $\psi_d$ is given in Table~\ref{tab:smallpsi}.
Likewise, an expansion in field around $H_{\rm sw}$ yields
\begin{equation}
  \label{eq:P_vs_H_MD}
	P(m>0)\Bigl|_t \Bigr. \approx \frac{1}{2} \left[ 1 +
		{\rm erf} \left(
		\frac{H - H_{\rm sw}}{L^{-d/2} \Delta_H}
		\right) \right] \ ,
\end  {equation}
where
\begin{equation}
  \label{eq:Delta_H}
	\Delta_H = \left( \tau \nu |H_{\rm sw}| \right)^{d/2}
		\left\{	\frac{d+K}{|H_{\rm sw}|} -
			|H_{\rm sw}|^{-d}
			\left[ (1-d) \, \Xi_0
                             + (3-d) \, \Xi_1 H_{\rm sw}^2 \right]
		\right\}^{-1} \psi_d^{-1} \ .
\end  {equation}
Here $H_{\rm sw}$ is the switching field corresponding to the
time $\tau$.

Valuable dynamical information can be found by fitting the MC data
to Eqs.~(\ref{eq:P_vs_t_MD}) and (\ref{eq:P_vs_H_MD}). From the
fit to Eq.~(\ref{eq:P_vs_t_MD}) shown in
Fig.~\ref{fig:P_vs_t}(c) we can find $\nu$ and $\tau$
at fixed $H$ directly (Tab.~\ref{tab:MDFits}).
In order to find $\nu$ from Eq.~(\ref{eq:P_vs_H_MD}) it is
first necessary to find $\Xi_1$.  We therefore fit
MC estimates of $\tau$ to Eqs.~(\ref{eq:NucRate}) and
(\ref{eq:MDLife}) to find
$B$ and $\Xi_1$~\cite{MDBXi1}.  The resulting estimates for $\nu$ are
consistent with the estimate from Eq.~(\ref{eq:P_vs_t_MD})
(Tab.~\ref{tab:MDFits}).  Note that for the range of fields
shown in Fig.~\ref{fig:P_vs_H}(c) the radial growth velocity
is less than one lattice constant per MCSS, which is to be
expected from a single-spin-flip dynamic such as we are using.

Knowledge of $B$, $\Xi_1$, and $\nu$ also allows us to find
an alternative estimate for the Dynamic Spinodal.
The crossover between the SD and MD regions can be expected to
occur where the mean waiting time for a droplet to nucleate is
approximately equal to the time required for a nucleated
droplet to grow to volume $L^d/2$.
This leads us to
\begin{equation}
  \label{eq:HDynEqual}
	\left[ L^d I (T,H) \right]^{-1} = (\nu |H|)^{-1}
	\left[ (2 \Omega)^{-1/d} L - R_c \right] \ ,
\end  {equation}
where $R_c$ is the critical droplet radius given in
Eq.~(\ref{eq:Rc}).
As shown by the dot-dot-dashed curve
in Fig.~\ref{fig:Hs_vs_L}(c), this estimate for
the Dynamic Spinodal is qualitatively similar to
the estimate given by Eq.~(\ref{eq:H_DSP}) but is quantitatively
different.   This is not surprising, since the Dynamic Spinodal
is a crossover region, and its precise location depends on the
manner in which it is defined.

It is also possible to approximately count the number of critical
droplets which nucleate during the lifetime of the metastable state.
By multiplying the nucleation rate by the volume of the system
and the lifetime [from Eq.~(\ref{eq:MDLife})]  we find
\begin{equation}
  \label{eq:NDrops}
  N_{\rm drop} = L^d I(T,H)
		\left[ \frac{\Omega \nu^d |H|^d}{(d+1)\ln 2}
		I(T,H) \right]^{- \frac{1}{d+1}} \ .
\end  {equation}
Figure~\ref{fig:MDSnapshot} shows a typical realization of
magnetization reversal in the MD region.  From Eq.~(\ref{eq:Rc}),
at $T \! = \! 0.8 T_c$ and $H \! = \! -0.2J$ a critical droplet
consists of approximately 48 overturned spins.
According to Eq.~(\ref{eq:NDrops}), approximately 18 critical
droplets should have nucleated in the system shown in
Fig.~\ref{fig:MDSnapshot} by the time $m \! = \! 0$,
which is in rough agreement with visual inspection of the figure.
However, it is difficult to confirm this number with precision
directly from the ``snapshots'' since
	  i) the droplets may partially or completely overlap,
	 ii) the droplets may have irregular shapes,
	iii) there are many subcritical droplets, and
	 iv) only a few times are displayed.

As a cautionary note, we analyze data taken in the SF region
using equations appropriate for the MD region.
As Fig.~\ref{fig:P_vs_t}(d) and Fig.~\ref{fig:P_vs_H}(d) show,
Eqs.~(\ref{eq:P_vs_t_MD}) and (\ref{eq:P_vs_H_MD}) can be fitted
to the MC data for $P(m \! > \! 0)$ quite well.  This is
because in the deterministic region, the observed magnetization
is essentially an average over the magnetizations of regions
too distant to be correlated, so that by the Central Limit
Theorem it tends to a Gaussian distribution as
$L \! \rightarrow \! \infty$.  Interestingly, both the value
of $\nu$ obtained from the fit in Fig.~\ref{fig:P_vs_t}(d)
[$\nu \! = \! 0.875(9)$] and the value from the fit in
Fig.~\ref{fig:P_vs_H}(d) [$\nu \! = \! 0.82(2)$] are in fairly
good agreement with the values derived in the MD region
(Tab.~\ref{tab:MDFits}).  However, the resulting radial growth
velocities ($v \! \approx \! 3$ lattice constants per MCSS) are
too large to be meaningful in a single-spin-flip dynamic.
Furthermore, whereas Eq.~(\ref{eq:MDLife}) gives a good fit
for $\tau$ in the MD region, it fails when applied to the
entire deterministic region.  Since it will be extremely
difficult to determine whether the radial growth velocity is
reasonable in experimental systems (due to uncertainty about the
appropriate classical dynamic and attempt frequency),
droplet theory should be applied only if both
Eq.~(\ref{eq:MDLife}) correctly gives the field-dependence of
the lifetime and $H \! < \! H_{\rm MFSp}$.

\section{Discussion}
\typeout{Discussion}
\label{sec-discuss}

Due to the importance of magnetic recording technologies in
modern society, magnetic relaxation has been a subject of
study for many years.  An early mean-field theory for this
process was presented by N{\'e}el \cite{Neel49} and
Brown \cite{Brown59,Brown63} almost fifty years ago, and the
most popular current models remain mean-field \cite{Koester,Kneller}.
This is understandable, since mean-field models are much
simpler to solve than models with local dynamics that describe
spatial fluctuations.
However, a great deal of research has been
conducted on non-mean-field descriptions of metastability,
particularly for the simple prototype model of anisotropic
magnetism, the Ising model.
These studies make it possible to obtain a more detailed
understanding of magnetic relaxation.

In this paper we have discussed the magnetic relaxation of
single-domain magnetic particles in terms of the
droplet theory of metastable decay.  We have performed
Monte Carlo simulations of the metastable decay of two-dimensional
Ising systems, and shown that these simulations yield switching
fields which vary with system size in a manner which is qualitatively
similar to experiments conducted on real single-domain particles
(Fig.~\ref{fig:Hs_vs_L}).
We have also used droplet theory to determine the
analytical forms of the probability that the magnetization remains
unswitched, and fitted these forms to our simulation data
(Figs.~\ref{fig:P_vs_t} and \ref{fig:P_vs_H}).
The methods used in carrying out these fits are appropriate for use
with experimental data obtained from magnetic force microscopy.

Because of their two-dimensional nature, the systems we have
simulated with MC actually serve better as models of ultrathin
islands of magnetic material on a nonmagnetic substrate.
(For a review of experimental observations of the magnetic
states of ultrathin films, see Ref.~\cite{Allenspach94}.)
The symmetry of our lattice and the use of
Ising spins require the modeled material to
have strong perpendicular anisotropy.  Several such ultrathin
films have been observed to have critical exponents consistent
with the universality class of the two-dimensional Ising model
(see \eg\ Refs.~\cite{Qiu91,Kohlhepp92,YiLi92,Qiu94,Elmers94}).
It would be interesting to compare microscopic observations
of magnetization reversal in these films with the kinetic
Ising model described in this paper.

Because films have macroscopic lateral extents ($L \! \gg \! 1$),
it is difficult to study any kind of finite-size effects in
them.  One way to impose a finite size to a film is to
grow it epitaxially on a stepped surface vicinal to a simple facet
so that the width of the step is of a microscopically moderate
size (see \eg\ Ref.~\cite{Chuang94} and references therein).
Of particular interest is Fe(110) on W(110), since at submonolayer
coverages the iron can be grown on step edges of the tungsten
surface and since the resulting film belongs to the
two-dimensional Ising model universality class \cite{Elmers94}.
For such semi-infinite systems Monte Carlo simulations such as
have been carried out here are not possible.
Sophisticated techniques in statistical mechanics involving
transfer matrices have, however, been successfully applied to
the two-dimensional Ising model in strip
geometry \cite{Guenther93,Guenther94} and to
a quasi-one-dimensional Ising model \cite{Gorman94}.
In both cases, it was found that concepts from droplet theory
are useful in describing the metastable decay.

It is important to understand the relaxation process in
the multi-droplet region clearly, both because it is relevant to
ultra-thin films and some single-domain particles and because
of controversies which have arisen over the mechanism of
magnetic relaxation \cite{AharoniMPFP}.  First of all,
we re-emphasize that droplets are distinct from domains
in that they are non-equilibrium entities.
Second, our model systems do not include defects or
impurities to serve as nuclei.  In fact, by using periodic
boundary conditions and the initial condition $m(0) \! = \! +1,$
all sites are completely equivalent; this symmetry is
spontaneously broken only by local thermal excitations.
Third, our model system is not a spin glass even though the
time-dependence of the magnetization may occur as a
stretched-exponential function
[Eqs.~(\ref{eq:app-Avrami}) and (\ref{eq:MDphi})].
Finally, the simulations we have carried out
are for an individual particle with a well-defined orientation.
This distinguishes it from some treatments of magnetic viscosity,
\eg\ Ref.~\cite{ChantrellMPFP}.
This last point also emphasizes the importance of experiments
like MFM which resolve the magnetic state
of isolated individual ferromagnetic particles.

\acknowledgements

The authors wish to thank S.~von~Molnar, D.~M.~Lind,
J.~W.~Harrell, W.~D.~Doyle,
and B.~M.~Gorman for useful discussions
and for comments on the manuscript.
This research was supported in part by the Florida State University
Center for Materials Research and Technology, by the FSU
Supercomputer Computations Research Institute, which is partially
funded by the U.~S.\ Department of Energy through Contract No.\
DE-FC05-85ER25000,
and by the National Science Foundation through Grants
No.\ DMR-9013107 and DMR-9315969.


\appendix
\section*{}
\label{app_rand}

Avrami's Law
\cite{Kolmogorov37,JohnsMehl39,Avrami39,Avrami40,Avrami41}
gives the volume fraction of the metastable state
(or equivalently, the magnetization) for systems in which
droplets nucleate with a rate (per unit volume) $I$ and grow
without interacting, although they may overlap.
The time-dependent system magnetization $\langle m(t) \rangle$
is given by
\cite{Kolmogorov37,JohnsMehl39,Avrami39,Avrami40,Avrami41}
\begin{equation}
  \label{eq:app-Avrami}
	\langle m(t) \rangle \approx m_{\rm s}
		\left[ 2 \langle \phi(t) \rangle - 1 \right] \ ,
\end  {equation}
where
\begin{equation}
  \label{eq:MDphi}
	\langle \phi(t) \rangle =
	\exp \left[-\ln2 \left( \frac{t}{\tau}
		\right)^{d+1} \right]
\end  {equation}
is the volume fraction of the metastable phase
and $\tau$ is given by Eq.~(\ref{eq:MDLife}).
The factor $\ln 2$ is necessary to fulfill the requirement
$m (\tau) \! = \! 0$, in agreement with the definition of
$\tau$ as the first-passage time to $m \! = \! 0$
[Eq.~(\ref{eq-define_tau})].

The nucleation of droplets is random in both
space and time, so only the mean magnetization is given by
Eq.~(\ref{eq:app-Avrami}).
The variance in the magnetization is related to
the two-point correlation function,
which has been calculated by Sekimoto \cite{Sekimoto86} for
circular and spherical droplets.
Specifically, if the system is described by a two-valued field
$u({\bf r},t)$ where
\begin{equation}
  \label{eq:def_u}
u({\bf r},t) = \left\{ \begin{array}{ll}
			1 & \mbox{if the position ${\bf r}$ is
				within the metastable phase at
				time $t$,} \\
			0 & \mbox{otherwise,} \\
		       \end  {array} \right.
\end  {equation}
then the connected two-point correlation function
is given by \cite{Sekimoto86}
\begin{eqnarray}
 \label{eq:C2u}
	 \langle u({\bf x},t) u({\bf x + r},t) \rangle
	  & - & \langle u({\bf x},t) \rangle^2 \nonumber \\
	  &   & =  \left\{ \begin{array}{ll}
		\langle u({\bf 0},t) \rangle^2
		\left\{ \exp \left[ Iv^d t^{d+1}
		  \Psi_d \left( r / 2 v t \right)
		\right] - 1 \right\}, & r < 2 v t \\
		0,                    & r > 2 v t \\
                      \end  {array} \right.
\end  {eqnarray}
where $r \! \equiv \! |{\bf r}|$, and \cite{Sekimoto86}
\begin{mathletters}
\begin{eqnarray}
  \label{eq:Psi_d}
	\Psi_2 (y) & = & \frac{2}{3} \left[
		\cos^{-1} y - 2 y \sqrt{1 - y^2}
		+ y^3 \ln \left(
		\frac{1 + \sqrt{1 - y^2}}{y}
		\right) \right] \ ,                      \\
	\Psi_3 (y) & = & \frac{\pi}{3} (1-y)^3 (1+y) \ .
\end  {eqnarray}
\end  {mathletters}
Since $\langle \phi(t) \rangle$ is simply an average over $
u({\bf r},t)$, with
\begin{equation}
  \label{eq:phims}
	\langle \phi(t) \rangle = L^{-d} \int
		\langle u({\bf r},t) \rangle {\rm d}{\bf r} \ ,
\end  {equation}
the variance of the magnetization is given by
\begin{mathletters}
\begin{eqnarray}
 {\rm Var}[m(t)]
\label{eq:VarM2} & \approx &
 	4 m_{\rm s}^2 L^{-d} \int \left[
          \langle u({\bf 0},t) u({\bf r},t) \rangle
 	 -  \langle u({\bf 0},t) \rangle^2 \right]
 	    {\rm d}{\bf r}
 \\ \label{eq:VarM3} & = &
	4 m_{\rm s}^2 d \Omega L^{-d}
	\int_0^{2 v t} r^{d-1} \left[
         \langle u({\bf 0},t) u({\bf r},t) \rangle
	 -  \langle u({\bf 0},t) \rangle^2 \right]
	    {\rm d}r
 \nonumber \\ \label{eq:VarM4} & = &
	4 m_{\rm s}^2 d \Omega \langle \phi (t) \rangle^2
	\left( \frac{2 v t}{L} \right)^d
	\int_0^1 y^{d-1}
	\left\{ \exp \left[ Iv^d t^{d+1}
		  \Psi_d ( y )
		\right] - 1  \right\}
	    {\rm d}y
 \nonumber \\ \label{eq:VarM5} & = &
	4 m_{\rm s}^2 d \Omega \langle \phi (t) \rangle^2
	\left( \frac{2 v t}{L} \right)^d
	\nonumber \\ & &
	\times \left\{ - \frac{1}{d}
	+ \int_0^1 y^{d-1} \exp \left[ \frac{(d+1) \ln 2}{\Omega}
			\left( \frac{t}{\tau} \right)^{d+1}
	  		\Psi_d (y) \right] {\rm d}y
	\right\} \ .
\end  {eqnarray}
\end  {mathletters}
The total system magnetization at time $t$ is
thus seen to be essentially an average over on
the order of $(L/2vt)^d$ independent samples.
As a result of the Central Limit Theorem, for
sufficiently large $L$, the distribution in magnetization
is well approximated by a Gaussian distribution.
Introducing the standard deviation in the magnetization,
\begin{equation}
  \label{eq:standev}
	\sigma(t) \equiv \sqrt{ {\rm Var}[m(t)] }
\end  {equation}
(not to be confused with the surface tension
$\sigma_L$ which is discussed in the text), it is easy to
calculate the probability that the magnetization is greater than
zero:
\begin{eqnarray}
  \label{eq:PGauss1}
	P(m > 0)
	& \approx & \int_{0}^{\infty}
		\frac{1}{\sigma(t) \sqrt{2\pi }}
		\exp \left\{ -\frac{[\mu - \langle m(t) \rangle]^2}
				  {2 \sigma^2(t) }
		     \right\} {\rm d}\mu \nonumber \\
  \label{eq:PGauss3}
	& = & \frac{1}{2} \left[ 1 +
		{\rm erf} \left(
		\frac{\langle m(t) \rangle}{\sigma(t) \sqrt{2}}
		\right) \right] \ .
\end  {eqnarray}
This probability will differ from zero or one only if
$\langle m(t) \rangle / \sigma (t) \! \approx \! 0$.
Making a Taylor expansion in time for
$\langle m(t) \rangle / \sigma (t)$ we find
\begin{equation}
  \label{eq:app_P_vs_t_MD}
	P(m>0)\Bigl|_H \Bigr.   \approx   \frac{1}{2} \left[ 1 +
		{\rm erf} \left( \frac{t - \tau}{L^{-d/2}\Delta_t}
		\right) \right] \ ,
\end  {equation}
where $\Delta_t$ is given by
\begin{equation}
  \label{eq:app_Delta_t}
	\Delta_t = \frac{( \nu |H_{\rm sw}| )^{d/2} \tau^{(d+2)/2}}
			{(d+1)\psi_d}
\end  {equation}
where $\nu$ is given by Eq.~(\ref{eq:def_nu0}),
$\tau$ is given by Eq.~(\ref{eq:MDLife}) with
$H \! = \! H_{\rm sw}$,
and $\psi_d$ is the constant given by
\begin{equation}
  \label{eq:smallpsi}
	\psi_d \equiv 	\ln2 \left(
			2^d d \Omega \left\{ - \frac{1}{d} +
			\int_0^1 y^{d-1}
			\exp \left[(d+1) \ln 2 \Psi_d (y) / \Omega
	  		\right] {\rm d}y \right\}
			\right)^{-1/2} \ .
\end  {equation}
Table~\ref{tab:smallpsi} lists numerical evaluations of $\psi_d$.
Likewise, a Taylor expansion of
$\langle m(t) \rangle / \sigma (t)$
in field yields
\begin{equation}
  \label{eq:app_P_vs_H_MD}
	P(m>0) \Bigl|_t \Bigr. \approx \frac{1}{2} \left[ 1 +
		{\rm erf} \left(
		\frac{H - H_{\rm sw}}{ L^{-d/2} \Delta_H }
		\right) \right] \ ,
\end  {equation}
where
\begin{equation}
  \label{eq:app_Delta_H}
	\Delta_H = \left( \tau \nu |H_{\rm sw}| \right)^{d/2}
		\left\{	\frac{d+K}{|H_{\rm sw}|} -
			|H_{\rm sw}|^{-d}
			\left[ (1-d) \, \Xi_0
                             + (3-d) \, \Xi_1 H_{\rm sw}^2 \right]
		\right\}^{-1} \psi_d^{-1} \ .
\end  {equation}
Here $H_{\rm sw}$ is the switching field corresponding to the
time $\tau$.


\begin{figure}
	\caption{
    	 	 \label{fig:HsRoad}
		 The relationship between the applied field $H$
		 and system width $L$ for various fixed lifetimes
		 (solid curves) in a typical metastable
		 magnetic system.
		 Four regions are distinguished by
		 differing decay processes:
		    the Coexistence    region (CE),
		    the Single-Droplet region (SD),
		    the Multi-Droplet  region (MD), and
		    the Strong-Field   region (SF).
		 The CE and SD regions are separated by the
		 thermodynamic spinodal (dotted curve).
		 The SD and MD regions are separated by the
		 dynamic spinodal (dash-dotted curve).
		 The SF region is separated from the other
		 regions by the mean-field spinodal (dashed curve).
		}
\end  {figure}

\begin{figure}
	\caption{
    	 	 \label{fig:Hs_vs_L}
		 The relationship between the applied field $H$
		 and system width $L$ for various fixed lifetimes
		 (solid curves) and temperatures as calculated
		 through kinetic Ising model simulations.
		 The thermodynamic spinodal (dotted curve)
		 is calculated from
		 Eq.~(\protect\ref{eq:H_THSP}).
		 The mean-field spinodal is approximated by
		 $H_{1/2}$ (dash-dotted curve),
		 calculated from Eq.~(\protect\ref{eq:H_DSP}).
		 The mean-field spinodal (calculated from
		 Eq.~(\protect\ref{eq:H_MFSP})) is not shown,
		 since for each of the three temperatures it occurs
		 at a field much larger than the displayed range.
		 a) $k_{\rm B}T \! = \! 0.65 J.$
		 b) $k_{\rm B}T \! = \! 1.30 J.$
		 c) $k_{\rm B}T \! = \! 0.8 k_{\rm B}T_{\rm c}
			  \! \approx \! 1.81535 J.$
		 The estimate for the Dynamic Spinodal given
		 by Eq.~(\protect\ref{eq:HDynEqual}) is shown
		 by the dot-dot-dashed curve.
		}
\end  {figure}

\begin{figure}
	\caption{
		\label{fig:P_vs_t}
		 $P(m \! > \! 0)$ \vs\ the time $t$
		 in MCSS for the kinetic Ising model at a
		 temperature of $T \! = \! 0.8 T_c.$
		 Note the different time scales.
		 (a) $L \! = \! 6$  and $H \! = \! 0 J,$
		     in the coexistence region.
		 (b) $L \! = \! 20$  and $H \! = \! -0.105 J,$
		     in the single-droplet region.
		 (c) $L \! = \! 100$ and $H \! = \! -0.34725 J,$
		     in the multi-droplet region.  The solid curve
		     is a fit to Eq.~(\protect\ref{eq:P_vs_t_MD}).
		 (d) $L \! = \! 30$ and $H \! = \! -3.5 J,$
		     in the strong-field region.
		}
\end  {figure}

\begin{figure}
	\caption{
		\label{fig:P_vs_H}
		 $P(m \! > \! 0)$ \vs\ the applied field $H$
		 for a kinetic Ising system at a temperature
		 of $T \! = \! 0.8 T_c.$
		 (a) $\tau \! = \! 4670$ MCSS and
		     $L \! = \! 6$,
		     in the coexistence region.
		     The horizontal line is $(1+e^{-2})/2$.
		 (b) $\tau \! = \! 914$ MCSS and
		     $L \! = \! 10$,
		     in the single-droplet region.  The solid
		     curve is a least-squares fit of
		     Eq.~(\protect\ref{eq:P_SD})
		     to the MC data.
		     The inset shows the fitted curve over
		     a wider range of fields.
		 (c) $\tau \! = \! 40.7$ MCSS and
		     $L \! = \! 30$, 100, and 300,
		     in the multi-droplet region.
		     The solid curves are fits of
		     Eq.~(\protect\ref{eq:P_vs_H_MD})
		     to the MC data.
		     The dashed curve is the fit of
		     Eq.~(\protect\ref{eq:P_vs_H_MD})
		     to the MC data for $L \! = \! 100$,
		     with the width of the switching region
		     reduced by a factor of $(300/100)^{-d/2}$
		     to compare with the MC data for
		     $L \! = \! 300$.
		 (d) $L \! = \! 30$ and
		     $\tau \! = \! 0.8$ MCSS,
		     in the strong-field region.
		}
\end  {figure}

\begin{figure}
	\caption{
		\label{fig:MDSnapshot}
		Spin configurations showing the nucleation and
		growth of several droplets in a typical realization
		of magnetization switching in the MD region.
		Here $L \! = \! 120$, $H \! = \! -0.2 J$, and
		$T \! = \! 0.8 T_c$.
		The times shown are at
			a) $t \! = \!  50$ MCSS,
			b) $t \! = \! 114$ MCSS, and
			c) $t \! = \! 160$ MCSS.
		Grey squares are ``up'' spins and black
		squares are ``down'' spins.
		}
\end  {figure}

\newpage
\narrowtext
\begin{table}
  \caption{Constants used in
	Eqs.~(\protect\ref{eq:P_vs_t_MD})
	and  (\protect\ref{eq:P_vs_H_MD})
	to calculate the
	width of the switching region as a function of time
	and as a function of field, respectively.
	The derivation is given in the Appendix.}
  \label{tab:smallpsi}
  \begin{tabular}{ld}
        $d$ & $\psi_d$     \\
       \tableline
         2  &    0.5628545 \\
         3  &    0.5109029 \\
  \end{tabular}
\end{table}

\newpage
\narrowtext
\begin{table}
  \caption{Results of fits of
	Eqs.~(\protect\ref{eq:P_vs_t_MD})
	and  (\protect\ref{eq:P_vs_H_MD}) to
	MC data for $P(m \! > \! 0)$ at $T \! = \! 0.8 T_c$.
	Estimated quantities are given with their
	statistical uncertainties. }
  \label{tab:MDFits}
  \begin{tabular}{rcddd}
        $L$ & Method & $\tau$ [MCSS]
		     & $|H_{\rm sw}|/J$ & $\nu J$ [1/MCSS] \\
       \tableline
       30 & $P$ \vs\ $H$ & 40.7    & 0.34684(2)  & 0.96(1) \\
      100 & $P$ \vs\ $H$ & 40.7    & 0.347253(2) & 0.92(2) \\
      100 & $P$ \vs\ $t$ & 40.8(4) & 0.34725     & 0.97(1) \\
  \end{tabular}
\end{table}


\begin{thebibliography}{10}
\bibitem{MDBXi1}
For $L \! = \! 30$ and $T \! = \! 0.8T_c$, the fits yield
  $B \! = \! 0.018(2) J^{-K}$/MCSS and
  $\Xi_1 \! = \! 2.0(2) J^{d-3}$.
  For $L \! = \! 100$ and $T \! = \! 0.8T_c$, the fits yield
  $B \! = \! 0.028(4) J^{-K}$/MCSS and
  $\Xi_1 \! = \! 3.0(3) J^{d-3}$.
  The discrepency between these results and those from the
  single-droplet region \protect\cite{SDBXi1} are probably due to
  finite-size effects in the small systems.
\end{thebibliography}
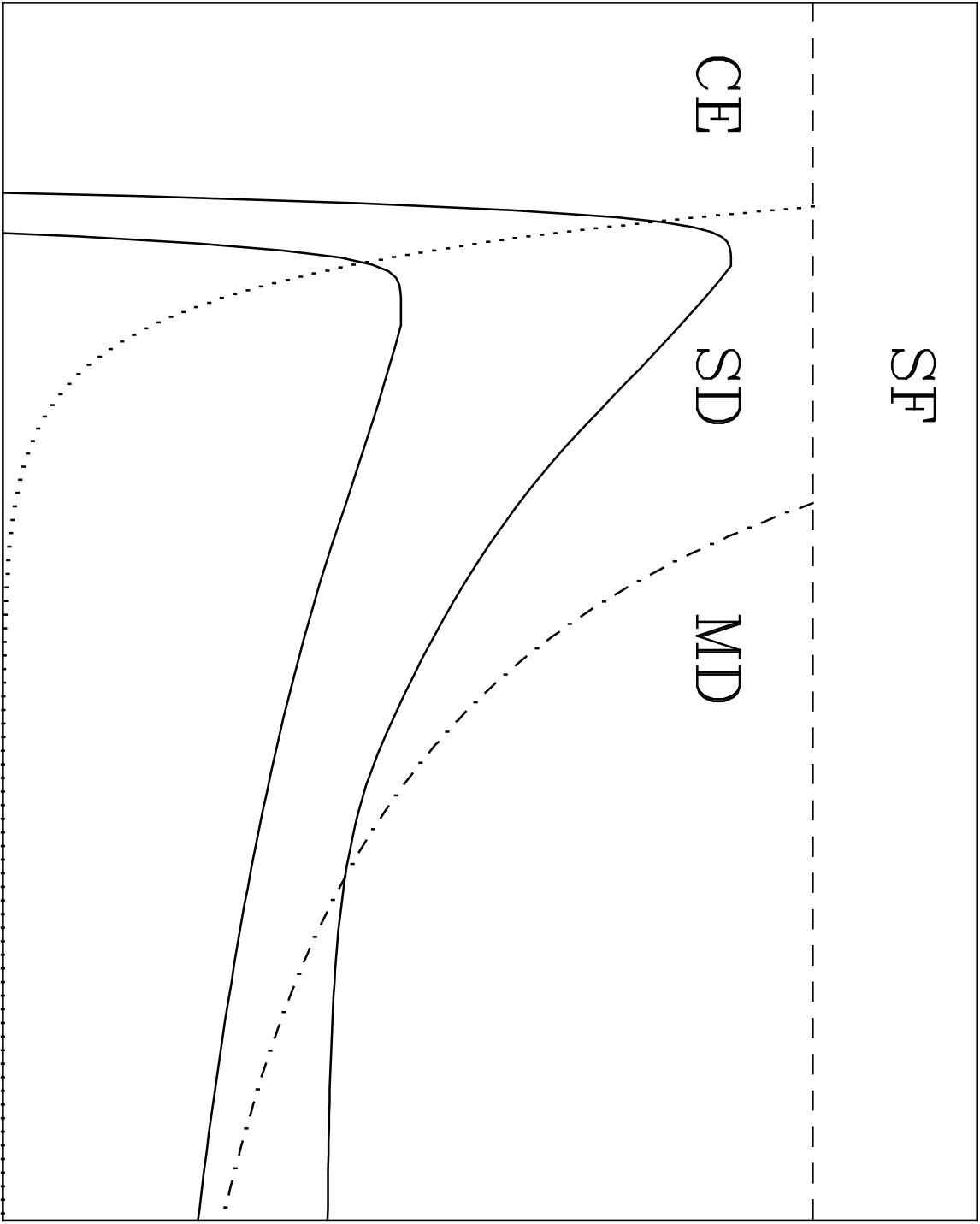
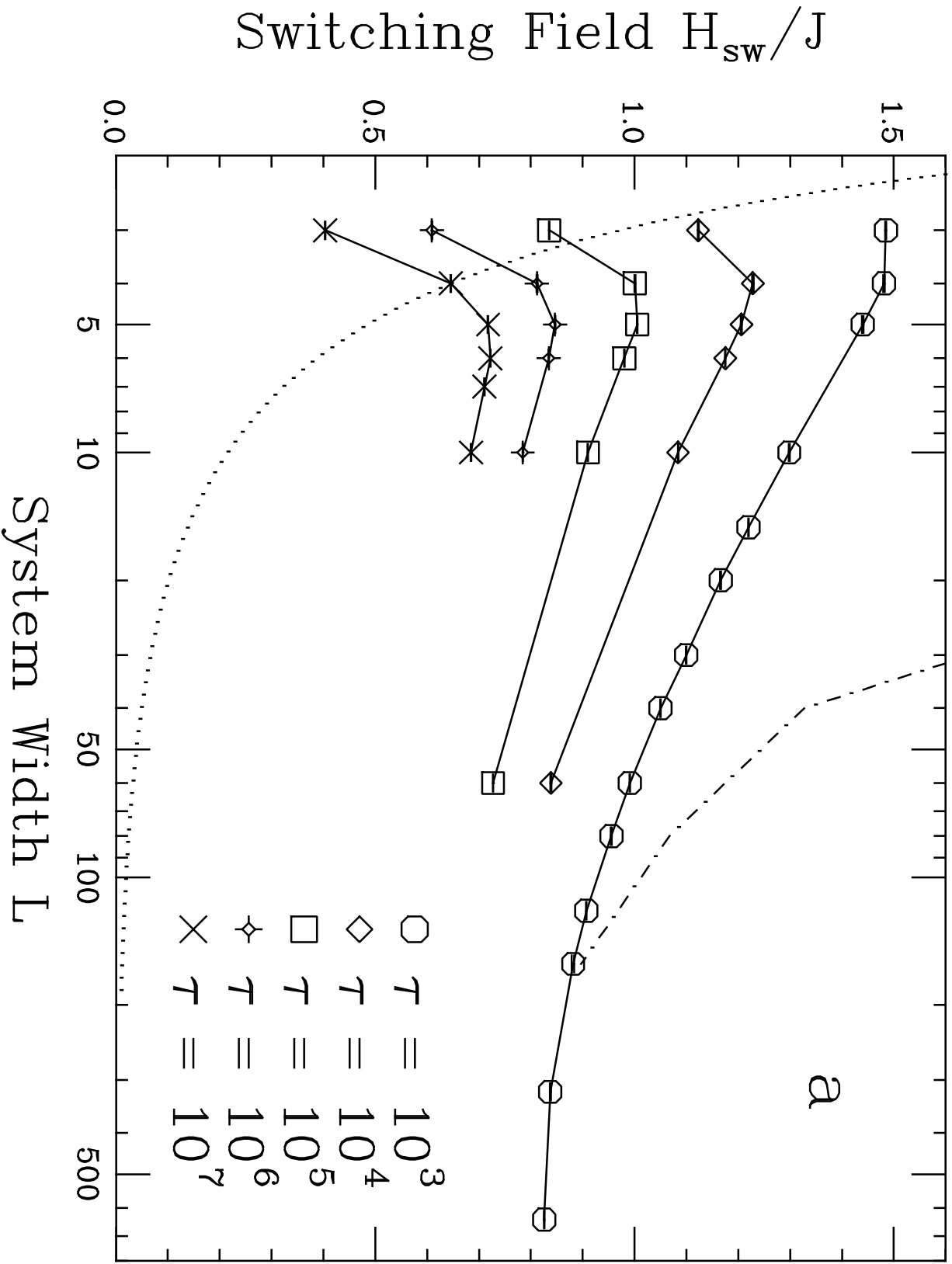
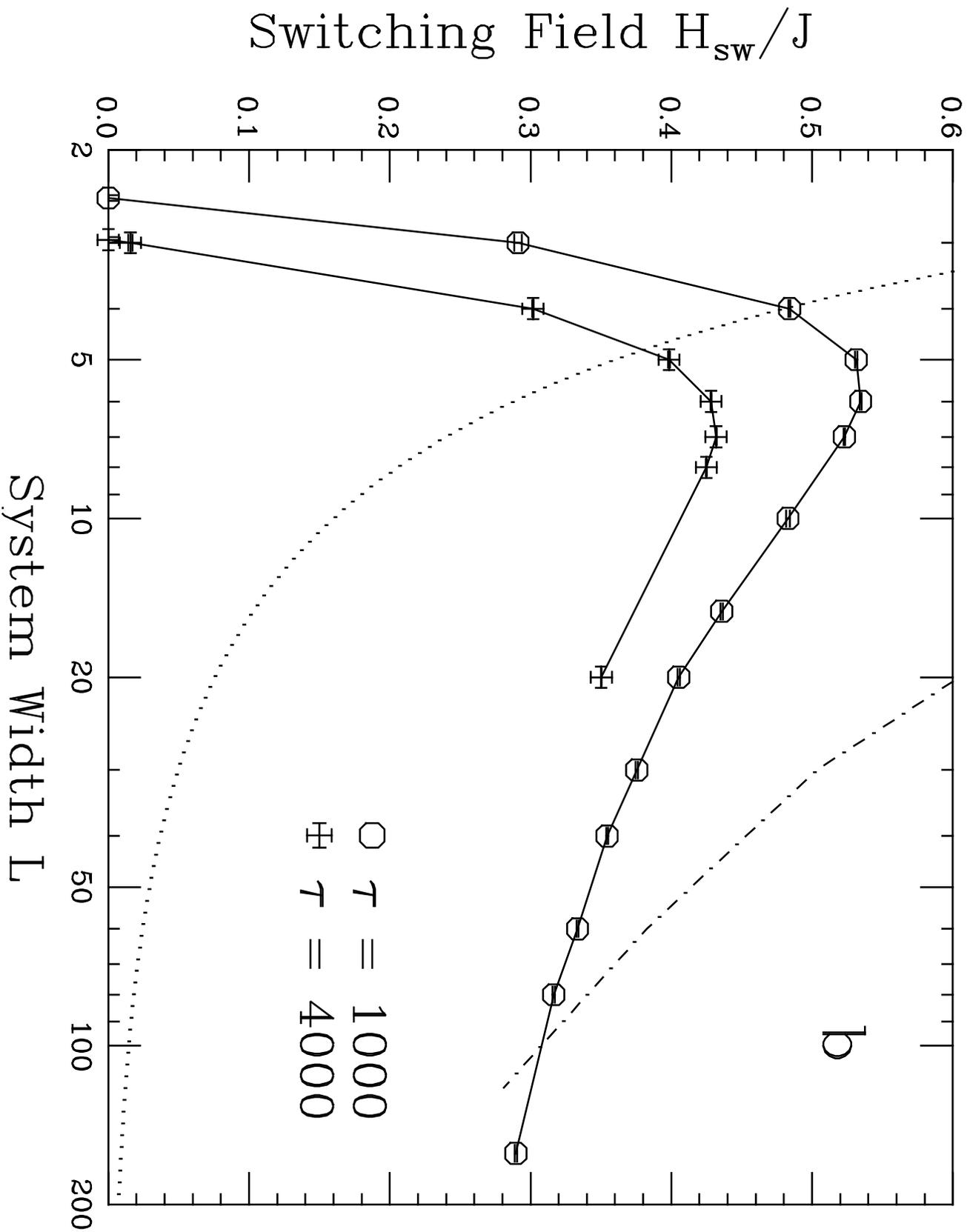
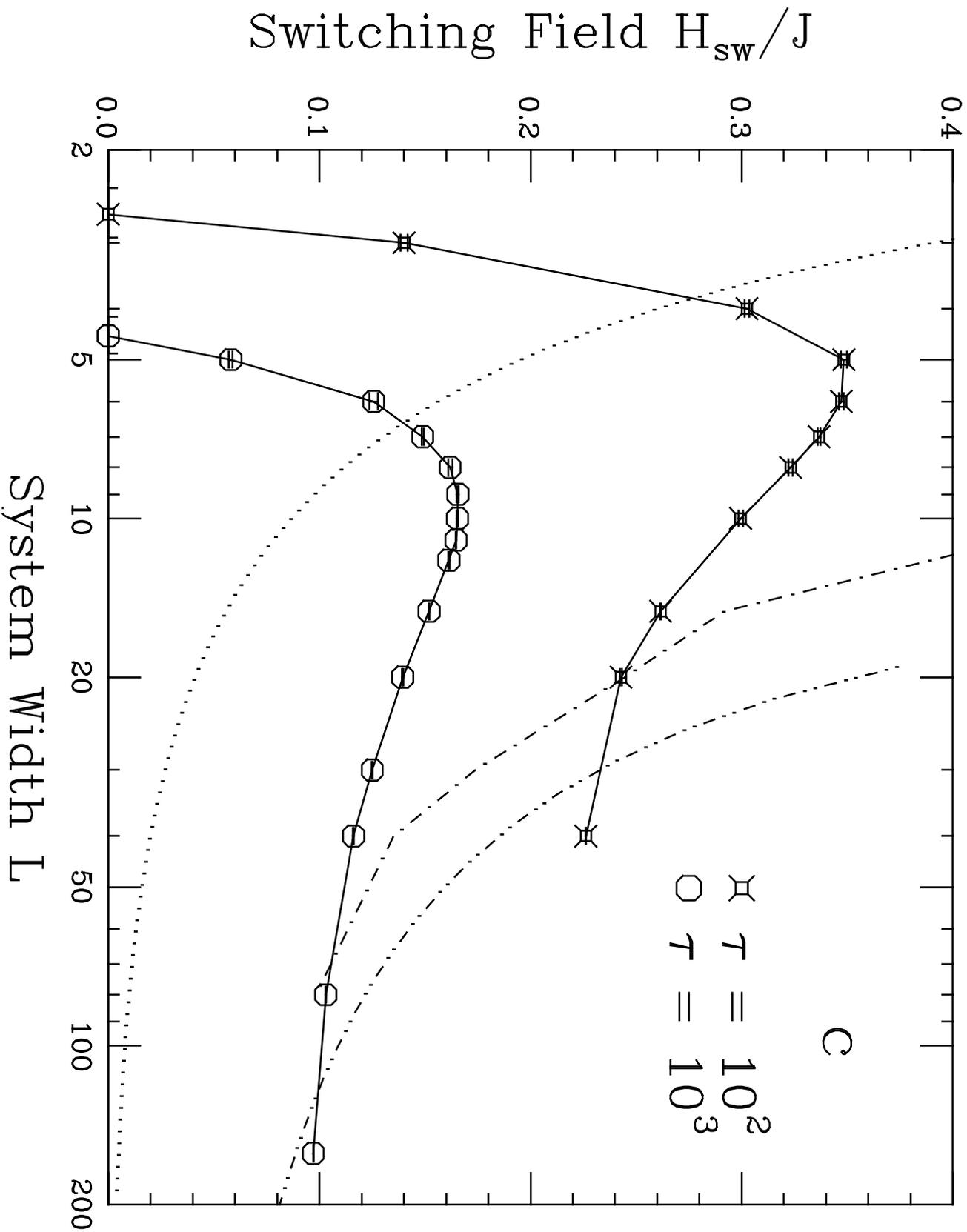
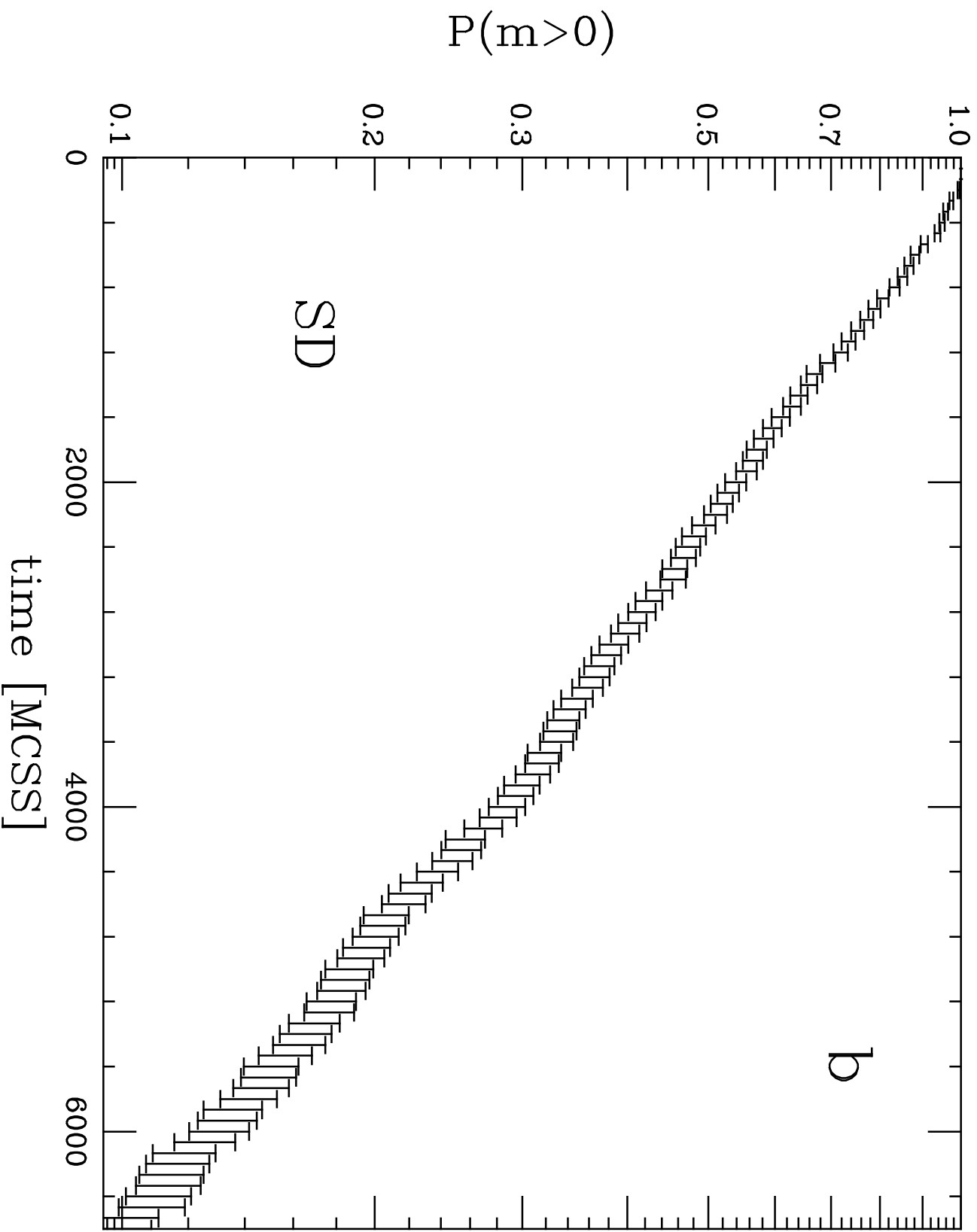
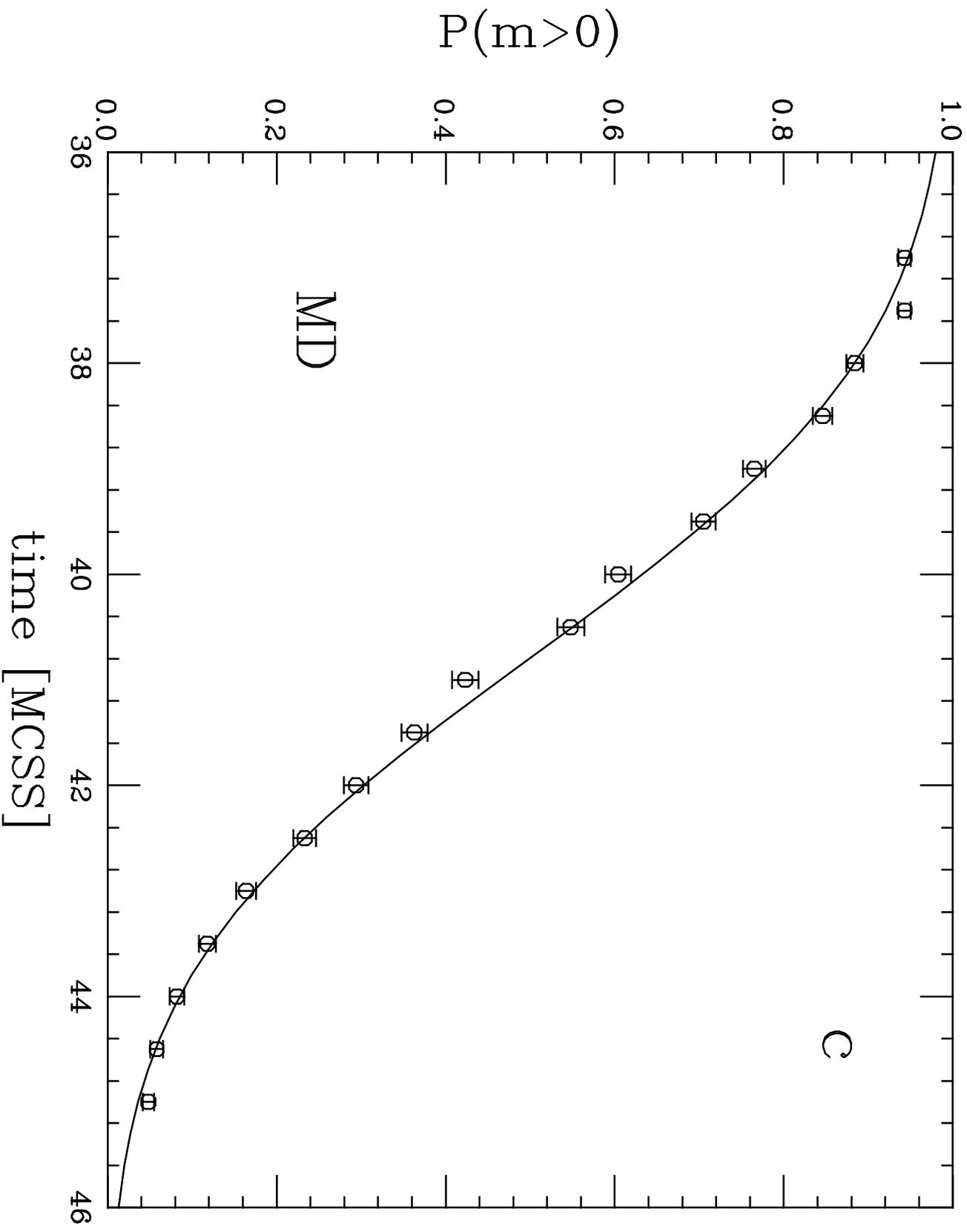
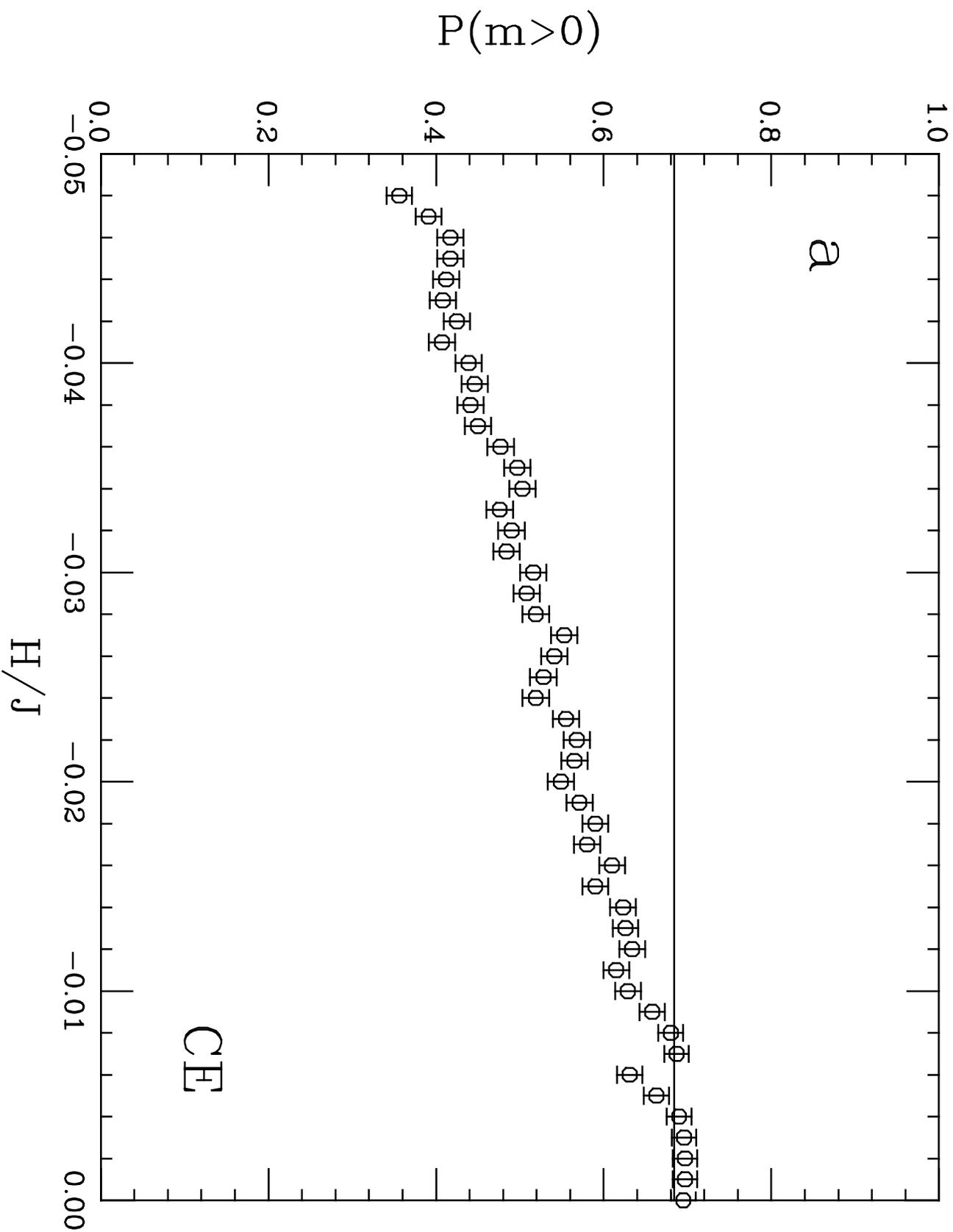
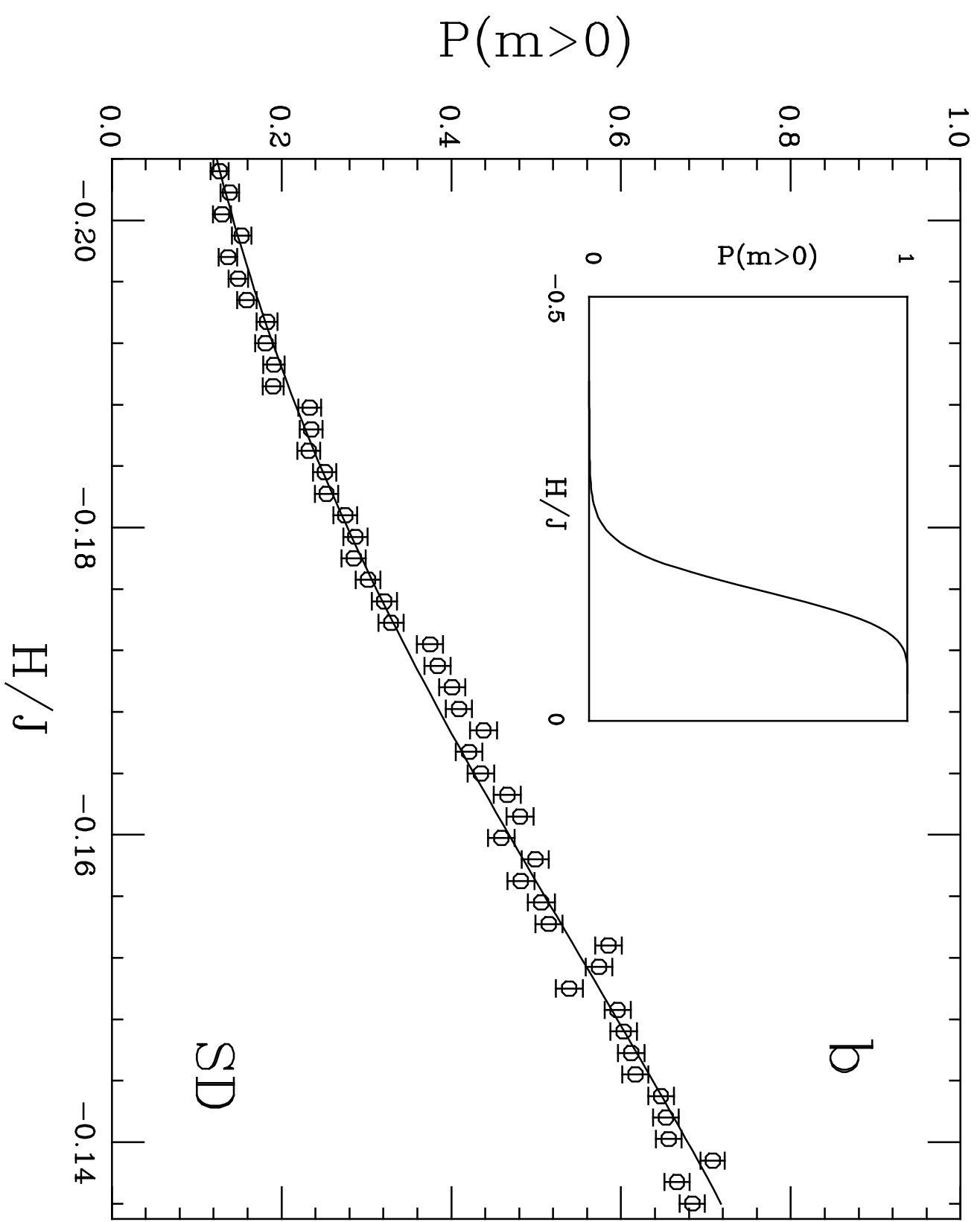
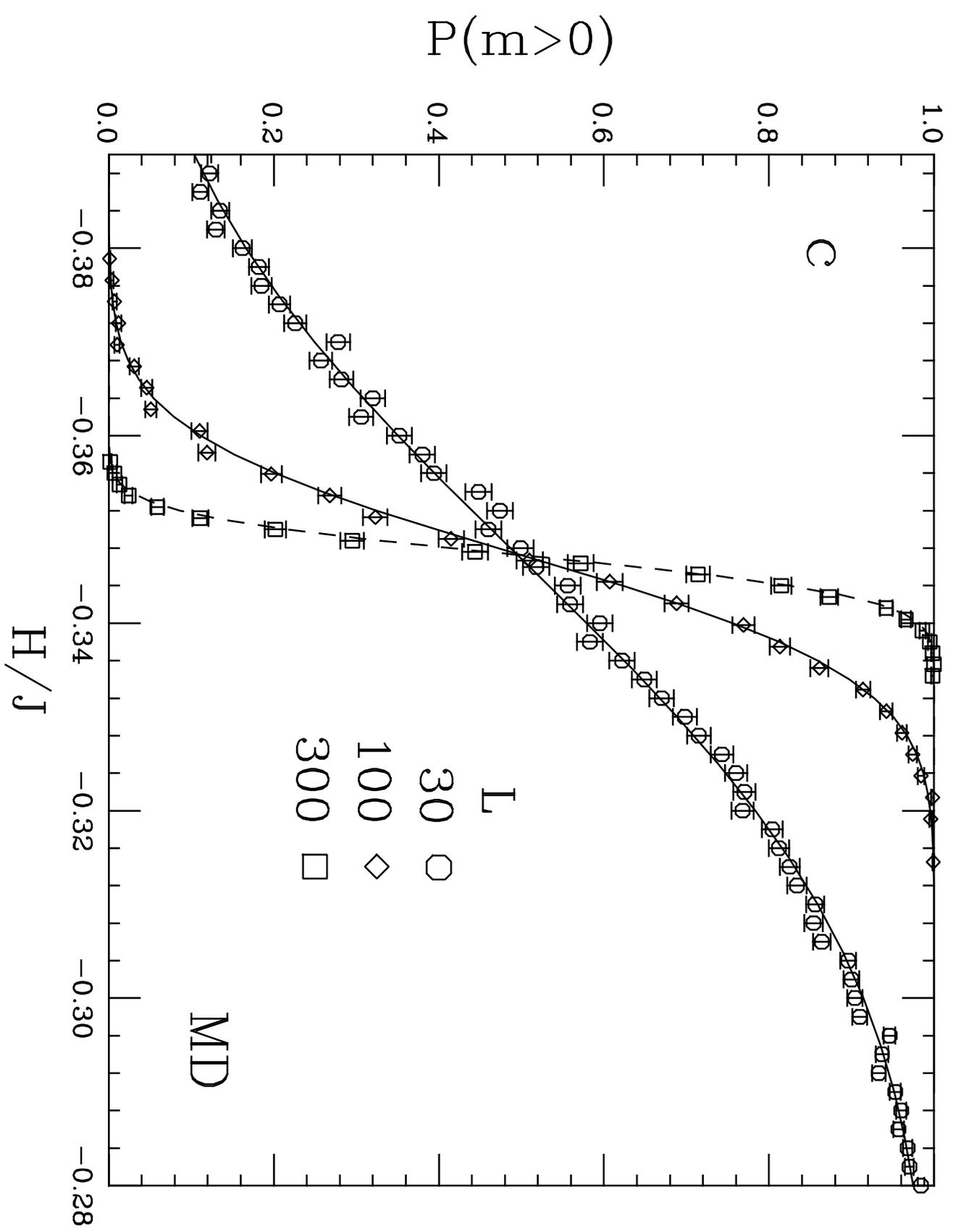
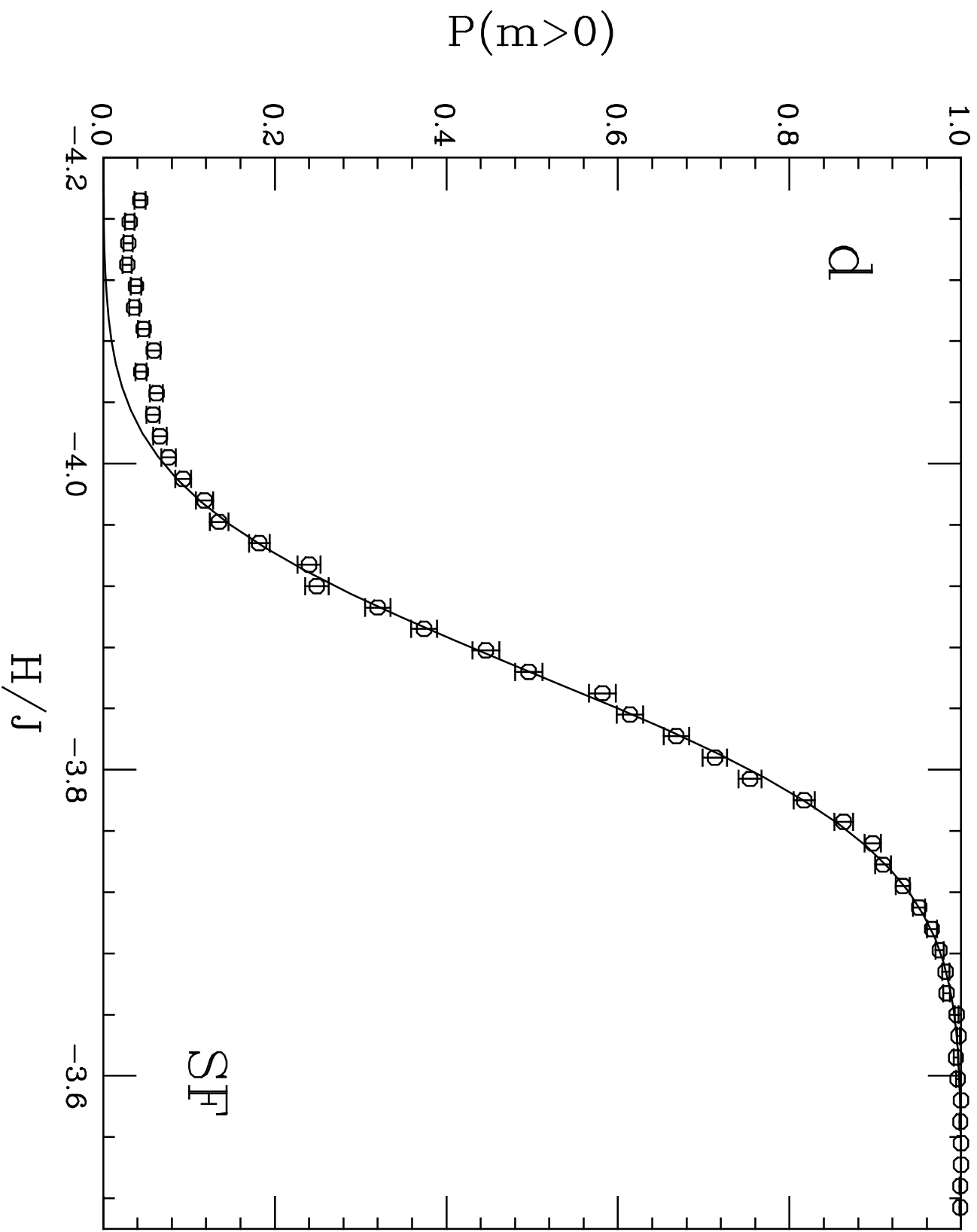
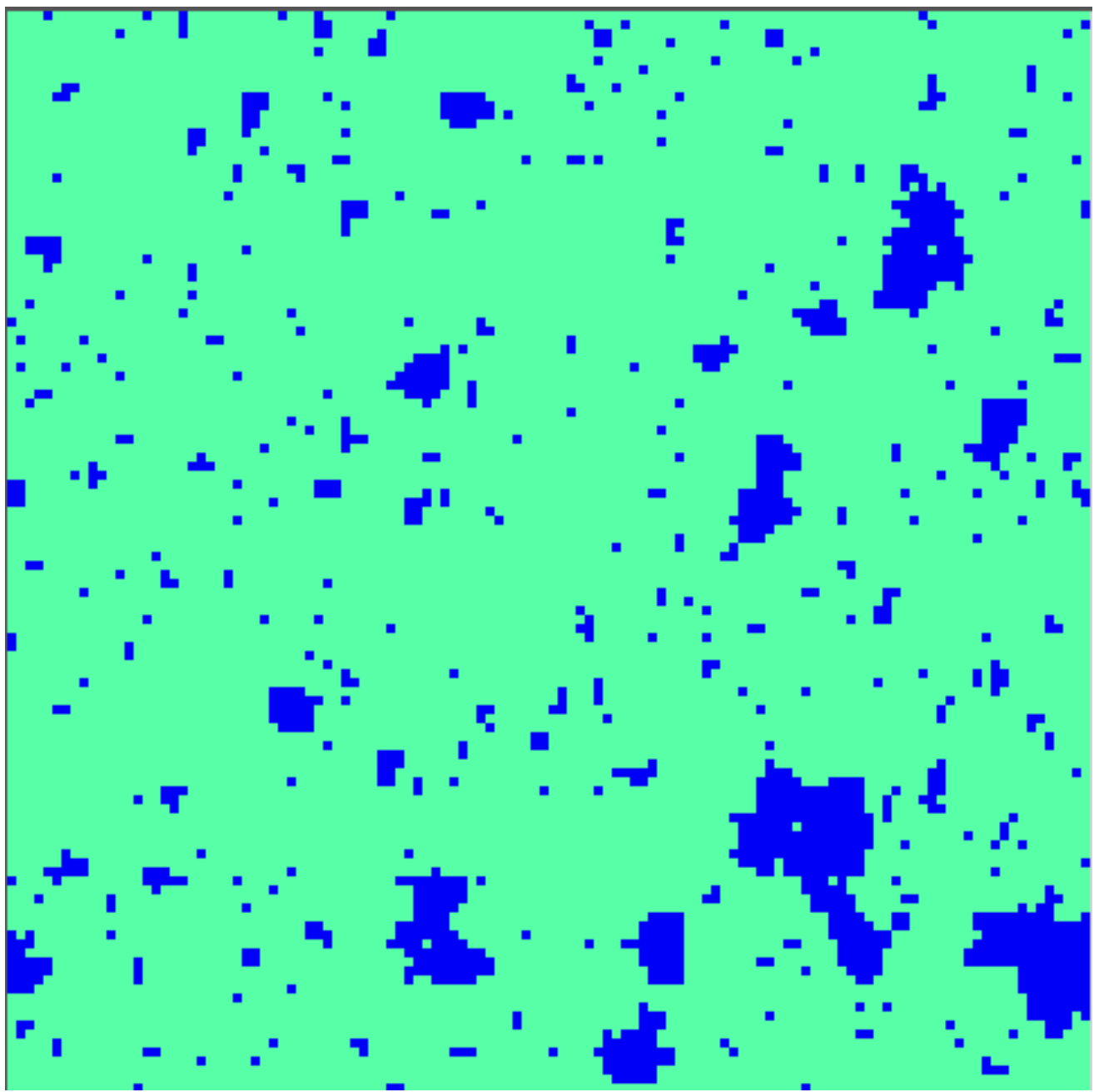
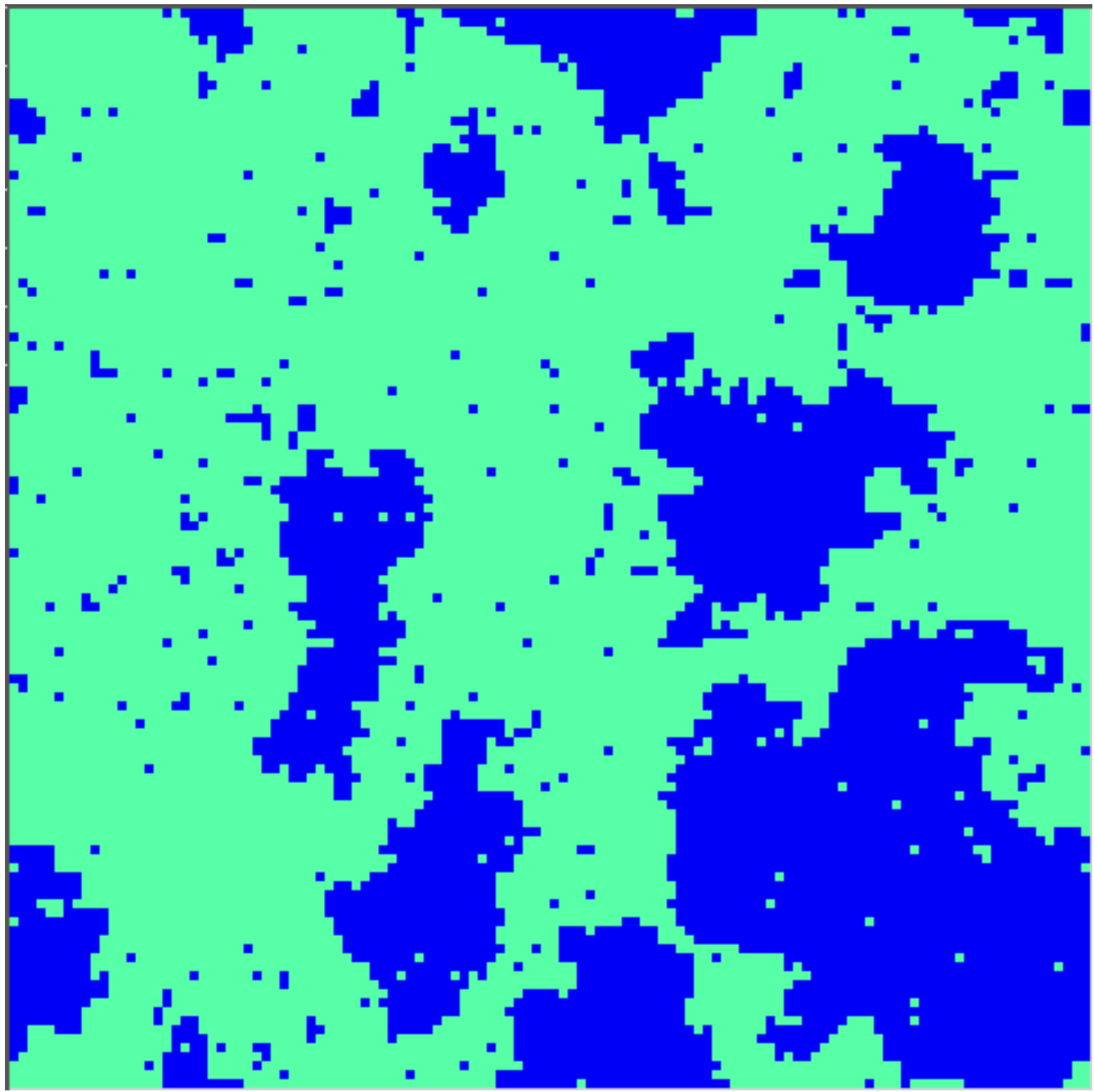
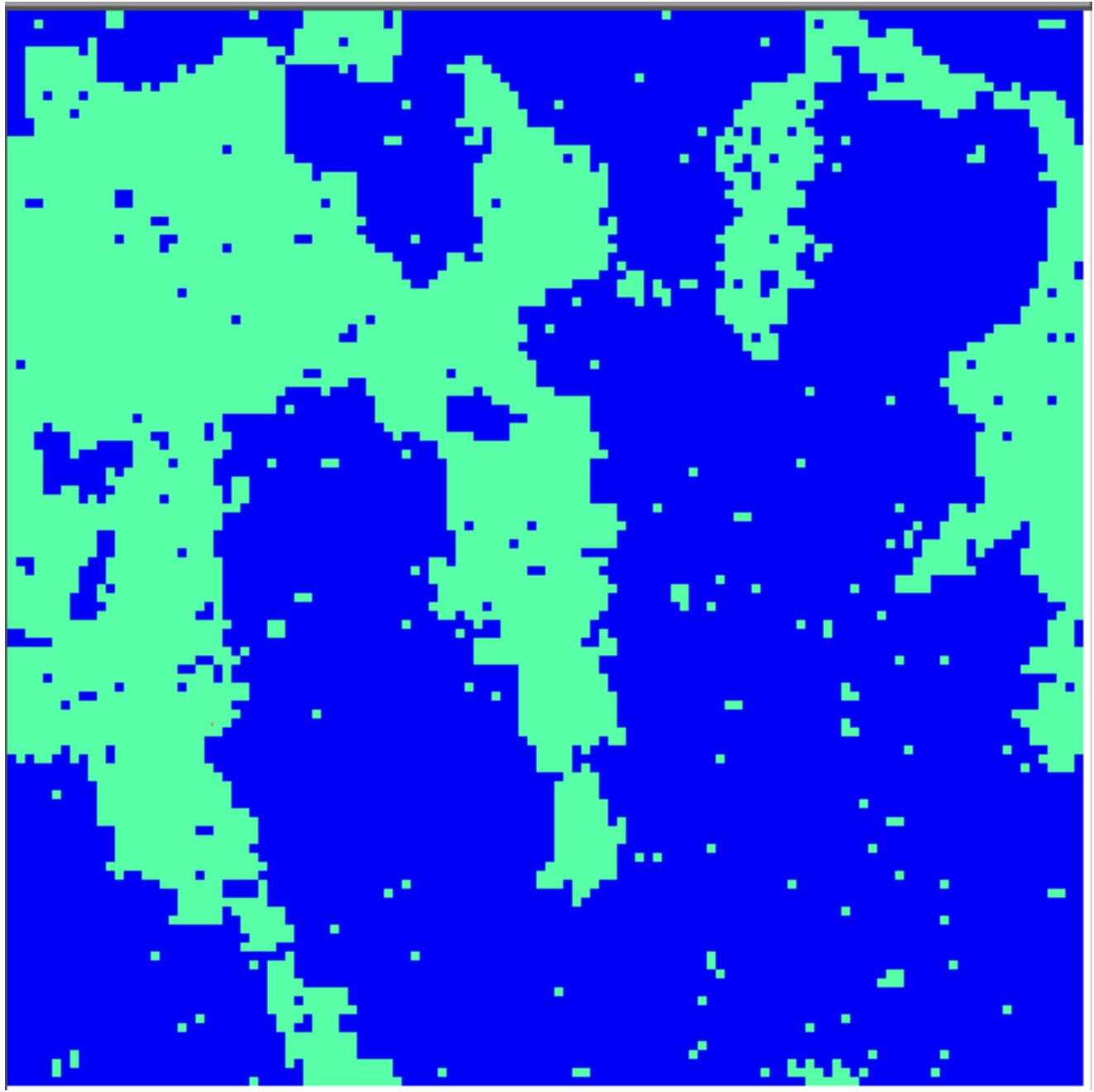
\end{document}